\documentclass[useAMS,usenatbib,usenatNishiNishibib,usegraphicx]{mn2e}
\usepackage{times}

\title[Galactic co-rotation gap in HI]{The corotation gap in the Galactic HI distribution}
\author[E. B. Am\^{o}res, J. R. D. L\'{e}pine and Yu. N. Mishurov]{E. B. Am\^{o}res$^{1,2}$
\thanks{EBA amores@astro.iag.usp.br(EBA)}, J. R. D. L\'{e}pine$^{1}$\thanks{jacques@astro.iag.usp.br(JRDL)}
 and Yu. N. Mishurov$^{3,4}$\thanks{mishurov@aaanet.ru}\\
$^{1}$ Instituto de Astronomia, Geof. e Ci\^encias Atmosf\'ericas
da USP, Cidade Universit\'aria 05508-900 S\~ao Paulo SP, Brazil \\
$^{2}$ SIM$/$IDL, Faculdade de Ci\^encias da Universidade de
Lisboa, Ed$.$ C8, Campo Grande, 1749$-$016, Lisboa, Portugal \\
$^{3}$ Space Research Department, Southern Federal University, 5
Zorge,Rostov-on-Don, 344090, Russia \\
$^{4}$ Special Astrophysical Observatory of Russian Academy of
Sciences,  Nizhnij Arkhyz, Karachaevo-Cherkesia, Russia}

\begin{document}

\date{Accepted XXXX December XX. Received XXXX December XX; in original form XXXX October XX}

\pagerange{\pageref{firstpage}--\pageref{lastpage}} \pubyear{2008}

\maketitle

\label{firstpage}

\begin{abstract}
We used the HI data from the LAB survey to map the ring-shaped gap in
HI density which lies slightly outside the solar circle. Adopting
$R_0$= 7.5 kpc, we find and average radius of 8.3 kpc and an
average gap width of 0.8 kpc. The characteristics of the HI gap correspond
closely to the  expected ones, as predicted by theory and by numerical
simulations of the gas flow near the corotation resonance.

 \end{abstract}

\begin{keywords}
Galaxy: structure -- Galaxy: kinematics and dynamics -- Galaxy: disc
-- Galaxy: general --  galaxies: spiral
\end{keywords}

\section{Introduction}

Since the first HI observations the existence of a ring-like minimum
in the gas density distribution in the Galaxy was noticed (eg. Kerr 1969a,
and Burton 1976). In these works, the authors found a gas
density deficiency at radius about $R = 11$ kpc, assuming $R_0 =$ 10 kpc,
but did not map it in detail nor proposed any explanation for it.  This
radius would be about $R = 8.3$ kpc, if we assume  $R_0 =$ 7.5 kpc. It
is tempting to associate this gap with the corotation radius ($R_c$), i.e.,
the radius where the rotation velocity of the spiral pattern coincides
with the rotation curve of the gas and stars. This association is suggested
on one hand, by the fact that it was shown, based
on different arguments or methods, that corotation is close to the orbit
of the Sun (Marochnik et al. 1972, Cr\'{e}z\'{e} \& Mennessier, 1973,
Amaral \& L\'{e}pine - hereafter AL-, 1997, Mishurov \& Zenina, 1999a,
 1999b, L\'{e}pine et al., 2001, among others).
More recently, Dias \& L\'{e}pine (2005, hereafter DL), using the
sample of open clusters, determined $R_c$ by measuring the spiral pattern speed
$\Omega_p$. The authors performed tests with different rotation curves
and $R_0$ values, obtaining $R_c$ = 8.1$\pm$0.6 for $R_0 =7.5$ kpc,
and argued that this determination is almost independent on
the rotation curve used.

The works mentioned above do not all adopt  the same galactic parameters,
but they are still in agreement, with values of $\Omega_p$ proportional to
V$_0$ so that $R_c$ results always close to $R_0$. However,
there are also in the literature many determinations of $\Omega_p$ giving
quite different values, and even some papers  proposing the existence of
multiple pattern speeds. Although the value of $\Omega_p$ is not the  focus
of the present paper, we will next comment a few of those results.  Naoz and
Shaviv (2007, hereafter NS), based on a similar analysis of the same data
used by DL, argued that the spiral arms situated in the solar vicinity
have different velocities, and  reached  the  surprising conclusion that
the Carina arm (which is usually considered as a normal arm, well fitted
by a logarithmic spiral) is made of two   superimposed components with
different  rotation speeds.  What is the origin of the contradiction
between the results of DL and NS?  In their analysis, DL used two
different methods to retrieve the birthplaces of the clusters. One is a
simplified analysis which assumes that the clusters move  in pure circular
orbits with constant velocity $\Omega(r)$; the second method performs
numerical integration of the orbits,  taking into account the observed
space velocities (radial velocities and proper motions)  of the clusters.
NS performed only the simplified analysis, with no integration of the
orbits, and this could explain the discrepancy. In a more recent paper
L\'{e}pine  et al. (2008) showed that the clusters have typical birth-time
velocities of the order of 10 kms$^{-1}$ with respect  to their local standard
of rest (the velocity of the rotation curve at the birthplace). This means
that the epicycle motion cannot be ignored  for young objects and
the constant velocity approximation is not satisfactory for precise
measurements. The anomalous velocities of the young stars in the Carina
arm is known (Humpreys \& Kerr, 1974). Other measurements tend to
favour DL{'}s results. Fernandez at al. (2001) analyzed the spiral
structure of the Galaxy using samples of OB stars and of Cepheids
and estimated $\Omega_p$ to be about 30 kms$^{-1}$kpc$^{-1}$,  reaching
the conclusion that corotation is near the solar orbit. Examining  the
details of their analysis, one sees that the OB stars have large errors
in distances (see Figure 1 of Torra et al., 2000, in which the sample is
described) while the sample of Cepheids is much better in this aspect.
Fernandez at al.  mention  that the   Cepheids  of their sample are
mostly situated in the Carina-Sagittarius arm. These stars have typical
ages about 10$^8$ years, being older than the OB stars, which minimizes
the problem of initial velocity.  The values of $\Omega_p$
that they obtain with their model "D" (the model which correspond to the
smaller value of $R_0$) coincide, within the small errors quoted by the
authors,  with $\Omega_p$ obtained by DL (24 kms$^{-1}$kpc$^{-1}$).
Avedisova (1989) also paid special attention to the Carina-Sagittarius
arm with a precise method and concluded that $\Omega_p$ is 26.8 $\pm$
2.2 kms$^{-1}$kpc$^{-1}$, in agreement with DL and Fernandez at al.
 In contrast  with the  methods based
on direct observations of sample of stars or clusters,  gas dynamics
(eg.  Bissantz et al. 2003, Rodriguez-Fernandez and Combes, 2008) and
N-body  or  test-particles  simulations (eg.  Chakrabarty , 2007) are
not able to provide  precise  determination of  the pattern speed, as
the authors of such models recognize; they usually "adopt" spiral pattern
speeds in the range 20-30 kms$^{-1}$kpc$^{-1}$(30-40 kms$^{-1}$kpc$^{-1}$
in the case of Rodriguez-Fernandez and Combes). Minchev and Quillen (2008)
predicts that the large-scale velocity surveys which are presently
planned will be able to constrain the galactic parameters, including
the pattern speed, in near future.

On the other hand, there are theoretical arguments and numerical
simulations telling us that we should expect a minimum of gas density at
corotation. The effect of the co-rotation can be understood as follows:
the dynamics of the gas in the potential perturbation of the spiral arms is such
that it produces a net flow  towards the center inside the corotation
radius, and a net flow towards the external parts of the disk,
beyond corotation, resulting in pumping out the gas from the corotation
region. Lacey \& Fall (1985) proposed an analytical expression for the gas
flow velocity on both sides of the resonance.
 Numerical hydrodynamic simulations were performed by Mishurov
(2007) in 3D and L\'{e}pine et al. (2001) in 2D.

The renewed interest in the study of the HI circular gap is related to
several of its implications  that we are presently investigating.
Firstly, since the corotation resonance is one of the fundamental
parameters of spiral galaxies (eg. Canzian,1998), its location in
the Milky Way is a question of special interest.
The existence of the gap is important to confirm our understanding
of the gas dynamics associated with the spiral structure of our
Galaxy. The gap could also, in principle, permit us
to infer a minimum "age" of the spiral pattern, and contribute to
the debate on the transient nature or the spiral structure.
Possibly the gap, if it is confirmed, could explain the minimum
in the rotation curve at a slightly larger galactic radius
(see eg. a description of the minimum by L\'{e}pine et al., 2008).
Indeed, as the rotation curve in the solar vicinity is almost
completely explained by the matter contained in the disk (eg. L\'{e}pine \&
Leroy, 2000), in the presence of a gap, a Keplerian-like decrease
of the rotation curve till the end of the gap would be expected.
Furthermore, the gap and the radial flow of gas in opposite
directions inside and outside corotation could be of great
importance for the models of chemical evolution of the disk, aimed
to explain the fine details of the gradient of metallicity.
The gap in some way could turn less efficient the exchange
of metallicity of the interstellar medium between the regions
situated on each side.

As an example of such fine structure in the  gradient of metal
abundance, Andrievsky et al. (2004) in a study of the Cepheids
metalicity as a function of galactic radius
found an abrupt reduction in the metalicity  between
10 and 11 kpc, assuming $R_0 = 7.9$ kpc. A similar step-like
distribution was found by Twarog et al. (1997) in the study of
metallicity of open clusters.

In the present paper, we use a kinematic distance method to
estimate the distance from the Sun of the minima of HI density
along the line-of-sight, for a large number of equally spaced
longitudes, so as to produce a map of the location of the minima
in the galactic plane. The method is based on the hypothesis
of circular rotation of the gas. We make use of the recently
published data from LAB survey to perform systematic analysis
in the HI data which covers both hemispheres with uniform
calibration.

The paper is organized as follows. Section 2 presents the data
from the LAB survey used in the present work. Section 3 presents
our method of analysis of the HI spectra and of mapping the
regions with lowest HI density. Section 4 presents a discussion of
evidences of the existence of the ring-shaped gap from other tracers
and in section 5 a short discussion of the existence of similar
features in other galaxies. Some of the consequences for the spiral
pattern of our Galaxy are discussed in section 6. The conclusions
and final remarks are presented in section 7.

\section{The HI Data}

Recently Kalberla et al. (2005) published the
LAB\footnote{Leiden/Argentine/Bonn} survey which contains the
final data release of observations of 21-cm emission from Galactic
neutral hydrogen over the entire sky, merging the Leiden/Dwingeloo
Survey (LDS: Hartmann \& Burton 1997) of the sky north of
-30$^{o}$ with the Instituto Argentino de Radioastronomia Survey
(IAR: Arnal et al. 2000 and Bajaja et al. 2005) of the sky south
of -25$^{o}$. The angular resolution of the combined material is
HPBW $\sim$ 0.6$^{o}$. One of the improvements of these new data
with respect to previous surveys
is the introduction of corrections for the stray radiation.

The LAB survey has been extensively used in several applications as
pointed by Bajaja et al. (2005), Kalberla et al. (2005), Haud \&
Kalberla (2007), Kalberla \&  Haud (2006) among others.
In the present paper, we have employed the HI data from the LAB
survey. These data cover galactic longitudes from 0$^{o}$ to
360$^{o}$ and galactic latitudes from -90$^{o}$ to 90$^{o}$; for
both coordinates the interval is 0.5$^{o}$ and the velocity resolution
1 kms$^{-1}$. The spectra are presented in units of antenna temperature
versus velocity. The data are stored in 720 $(b,v)$ fits file maps at
longitude intervals stepped by 0.5$^{o}$.

\section{Analyzis of the spectra and discussion of the existence of the gap}

We analyzed the HI spectra of the whole galactic longitudes range
in steps of 0.5$^{o}$, with galactic latitudes  in steps of
one degree in the range $\pm 5^{o}$, plus the additional latitudes
$\pm 10^{o}$. In each spectrum we detected the velocity of the deep minima
which are present, simply by identifying the channel with the lowest value
of antenna temperature. We previously smoothed the spectra by replacing
the temperature in each channel by the average of 5 channels.

Figure 1 illustrates what we call a deep minimum. It shows the HI
spectra for six directions, all at b=$0.0^{o}$, except for the last frame.
The vertical lines represent the detected location of the minima.
For instance, the HI spectrum obtained at $\ell =$ 240.0$^{o}$, presents
with a clear gap at $v \sim$35 kms$^{-1}$; this gap is so prominent that
it seems to divide the spectrum in two independent regions. As a working
definition, we considered that a "gap" or a "deep minimum" must be a
minimum in the spectrum where the antenna temperature is lower than
15 \textit{K}, but is comprised between regions of the spectrum with
antenna temperatures at least 20 \textit{K}above the minimum. In this way we
only consider minima which are significant in terms of signal to noise
ratio and which are not at the edges of the HI distribution, where
the density becomes naturally low. We also took separately into
consideration the "not very deep" gaps, with minima in the range 15 to 30
\textit{K}, but also with edges reaching at least 20 \textit{K} higher
than the minimum. In some cases there are
more than one gap in a same spectrum. One example is the spectrum at
$\ell =$ 80.0$^{o}$, which shows two minima that are not very deep.

 Usually, for a given longitude,
a gap can be seen at the same velocity in the spectra obtained at
different latitudes, but it disappears at $b= \pm 10^{o}$. But it also
happens in a number of directions that the gap is not prominent at
b = $0.0^{o}$, but is easily seen at some other latitude
$\mid$ b$\mid \leq $5.0$^{o}$. For instance at longitude
$\ell =$ 340.0$^{o}$, there is no deep minimum at $b= 0^{o}$,
but an almost good minimum can be seen at $b= -1^{o}$ (last frame of Figure 1).
This one is not taken into account in our following statistics
because it does not fulfill the condition of having emission higher
that 20 above the minimum on both sides. And since we are willing to
illustrate in Figure 1 not only the best cases, the minimum
in the $\ell =$ 280.0$^{o}$ frame at 70 km/s is a deep one, but
the minimum which would give the best fit to the interpretation that
we present in next sections is the one at 20 km/s. However,
the minimum at 20 km/s is present at the same longitude but
other latitudes, like $b= 1^{o}$ and $b= -3^{o}$, so that
it is taken into account. All the  measurements that show
a deep minimum or a "not very deep"  minimum as we defined them
are considered in the following discussion.

\begin{figure*}
\centering \resizebox{16.0cm}{10.0cm}{\includegraphics{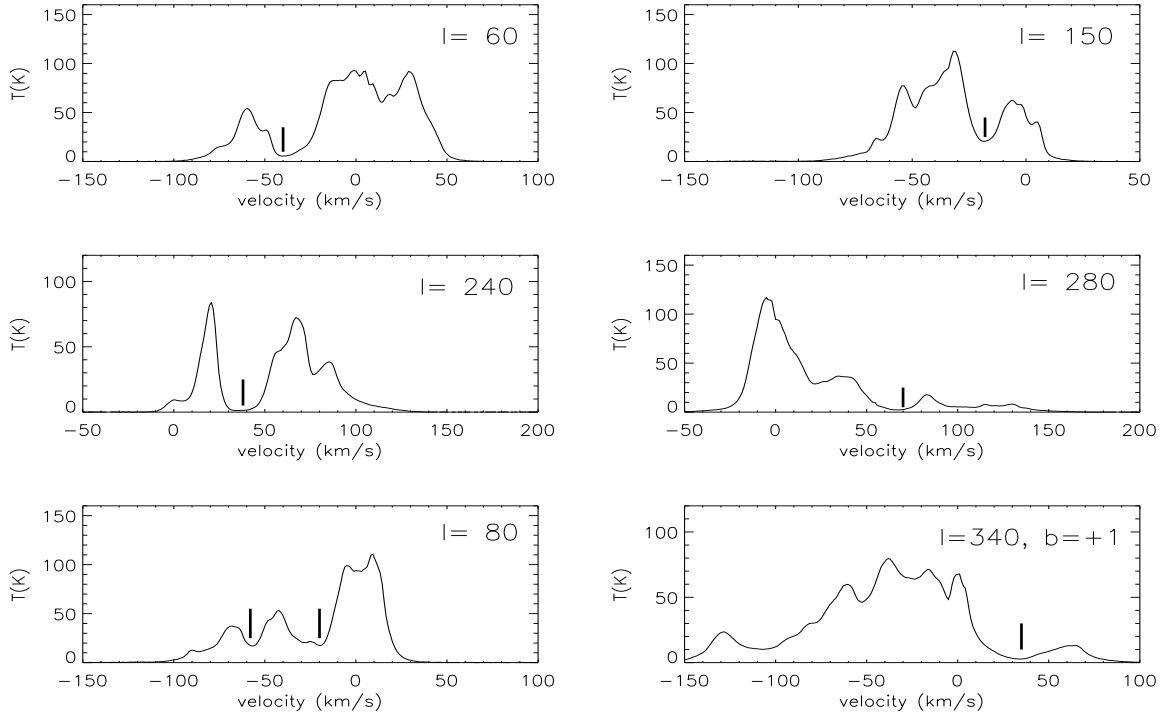}}
\caption{HI spectra for 6 different longitudes; the longitudes
are indicated in the figures, all at $b$ = 0$^{o}$ except for $\ell =$
340$^{o}$, which is at $b$ =1$^{o}$. The vertical thick lines
represent the location of the deep minima. The intensities are in
units of antenna temperatures (K).}
\end{figure*}

The result of the first step of analysis discussed above is a catalog
of velocities of deep minima and of not very deep minima as a function
of longitude. We present in Figure 2 the $\ell$-\textit{v} diagram of
these minima. The dashed line represent the locus in the $\ell$-\textit{v}
diagram of the points situated on a circle of radius 8.3 kpc around
the Galactic center, in the Galactic plane, as discussed later in
greater detail.

The line segments along the longitude axis at zero velocity represent
the "forbidden regions", or ranges of longitudes where we should not
see the effect of a ring-shaped gap in the HI distribution, even
if it is physically present. For instance, in the directions 0$^o$ and
180$^o$ the gas which rotates with the Galactic disk is supposed to
cross  the line-sight at right angles, and therefore, to contribute
 only to a narrow region around v = 0 kms$^{-1}$ in the spectra. The reason
 why the directions 90$^o$ and 270$^o$ are also partially forbidden
 is more subtle. This can be seen from one of the classical Oort formulae,
$$V_r = A d~ sin(2\ell)~~~~~~~~(1)$$
where $V_r$ is the radial velocity for an object situated at a distance
$d$ in the direction $\ell$, and $A$ is Oort\'~s constant. This expression
is valid for distances from the Sun that are not too large (let us say,
smaller than about 2 kpc). This expression tell us that the velocity
is zero also for 90$^o$ and 270$^o$. In other words, all the emission
from the gas situated up to about 2 kpc contributes to a narrow velocity
range in the spectrum, around v = 0 kms$^{-1}$. Gas clouds situated at larger
distances in those directions, however, contribute to non-zero
 velocities. If we exclude the "forbidden" regions, about $\pm 20^o$
 in the directions 0$^o$ and 180$^o$ and $\pm 10^o$ at 90$^o$ and
 270$^o$, and if we take into account the "not very deep" minima,
 then we can see that the gap is present along the dashed line
 in about 90\% of the permitted longitude range. In the next section
 we conclude that the ring-shaped gap has a radius about 8.3 kpc,
 which turns it more distant than 2 kpc in the directions 90$^o$ and
 270$^o$. We understand {\it a posteriori} why we can see the gap in those
 directions too.

 \begin{figure}
\includegraphics[width=84mm]{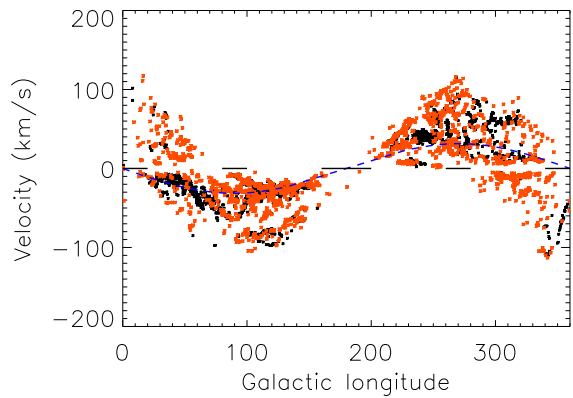}
\caption{The longitude-velocity diagram of the "deep" minima
and "not very deep" minima observed
in the spectra. The horizontal axis is galactic longitude in degrees,
and the vertical axis velocity in kms$^{-1}$. The deep minima are
shown in black, other gaps with minimum in the range 15-30 \textit{K}
(see text) are shown in orange. The blue dashed line
correspond to a circle of 8.3 kpc radius in the galactic plane. The
line segments along the longitude axis represent the longitude ranges
where the minima should not be seen even if the ring-shaped gap exists
along the whole circle.}
\label{figure2}
\end{figure}

 The question that we next address is  why the gap
 is not well seen in 100\% of the "permitted" longitude range
 (the whole longitude range excluding the directions close to 0$^o$or
 360$^o$, and 180$^o$). The minimum is not observed in a spectrum
 if there is a region along the line-of-sight containing gas with the
 same velocity of the gap. The velocity overlap is often only partial.
 We remark that when we are less rigorous in the definition
 of a gap, i.e., when we consider that the minimum can be larger than 15 K
 (the "not very deep minima") the longitude range covered is greater.
 For instance the gap is very clear $\ell = 150^o$ at the velocity $\sim$
 -18 kms$^{-1}$, where it is expected, but the antenna temperature
 at the minimum is 20 K (the spectra of selected directions can be
 easily obtained from the site of the LAB Survey
 \footnote {http://www.astro.uni-bonn.de/\~webaiub/english/tools\_labsurvey.php}).
At $\ell = 160^o$ there is still a minimum at v = -10 kms$^{-1}$,
but it is much less prominent, with  39 K at the bottom.
Similarly, at $\ell = 210^o$ there is a minimum at v = 20 kms$^{-1}$,
but the minimum is 39 K.

The more extended longitude range where the gap (as we defined it)
is only weakly observed (but still present), although it is a "permitted"
range, is 200-220$^o$. A possible explanation is the following.
 There is a spiral arm passing relatively close to the Sun (at about 2kpc
 in the direction of the anticenter; see eg. Russeil, 2003). Since the distance
 of the gap that we are investigating is about 1 kpc from the Sun
 towards the anticenter, the distance between the gap and the arm
 is of the order of 1 kpc. The slope of the curve of radial velocity
 as a function of distance along the line-of sight, for $\ell = 210^o$
 (in the middle of the longitude range that we are
 considering) tells us that a distance interval of  of 1 kpc
 produces a difference of velocity of the order of 10 kms$^{-1}$.
 On the other hand, the peaks seen in the spectra at any longitude
 never have width smaller than about 10 kms$^{-1}$. This is a minimum
 width of the HI emission associated with spiral arms, attributed
 to the turbulent motion of the gas. Therefore, it is not surprising that
 the minima in the spectra which are associated with the ring-like density gap
 are partially filled with emission which originates in a nearby region
 along the line-of-sight.

 \subsection{Detailed description of the gap}

 The kinematic distance of the gap from the Sun can be easily
estimated, as there is no distance ambiguity, since the gap is situated
outside the solar circle, where there is only one kinematic solution.
Furthermore, the distance obtained depends very little on the
adopted rotation curve, since the gap is close to the solar circle.
It must be remembered that if the gap were coincident with the
solar circle, its velocity as seen from the LSR would be zero
at all longitudes  independently of the rotation curve.

In the present work, we  used a relatively flat rotation
curve, conveniently fitted by exponentials and and a Gaussian
(units are km$^{-1}$ and kpc):

$$V =245\exp\left[-r/75.0 - (3.6/r)^2 \right] + 350\exp\left[-r/3.3 - 0.1/r \right]$$
$$-20\exp\left[-((r-8.8)/0.8)^2 \right]~~~~~~~~~~ (2)$$

The curve is presented in Figure 3, where the points represent
the CO data  obtained by Clemens (1985) corrected for
$R_{0}$ = 7.5 kpc and $V_{0}$ = 210 kms$^{-1}$. This curve is
close to that derived by Fich, Blitz \& Stark (1989) and the fitted
expression is similar to one previously used by our group
(eg L\'{e}pine et al. 2008)

\begin{figure}
\includegraphics[width=84mm]{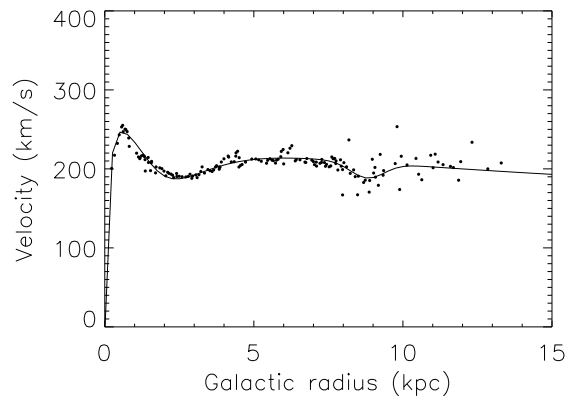}
\caption{The rotation curve of the Galaxy for $R_0$ = 7.5 kpc.
The line is a fit to the CO data, using the empirical expression
of Equation (2).}
\label{figure3}
\end{figure}

The distribution of the density minima in the galactic plane, derived from
their kinematic distances from the Sun, is shown in Figure 4.
The minima observed at different latitudes are all projected on
the galactic plane. The ring-shaped gap is circular and very clear.
It looks like the Cassini division in Saturn\'~s rings.
Although minor changes in the galactic radii of the gaps appear in
experiments with different rotation curves, the ring-shaped aspect
of their galactic distribution remains unchanged. The histogram
of the galactic radii is shown in Figure 5.
Most of the other structures not belonging to the ring have
the shape of spiral arms and can be understood as inter-arm gaps.
The prominent one at about 11 kpc to the left of the center
in the figure seems to be on the external side of the
Carina-Sagittarius arm. Taking into account the different galactic
scales adopted in the two works, it coincides with an inter-arm
region previously observed by Levine et al.(2006) at 13 kpc to the
left of the center. Figure 4 tells us that the ring-shaped
gap cannot be explained as a combination of inter-arm gaps,
nor it is a local structure only seen close to the Sun.

\begin{figure}
\includegraphics[width=100mm]{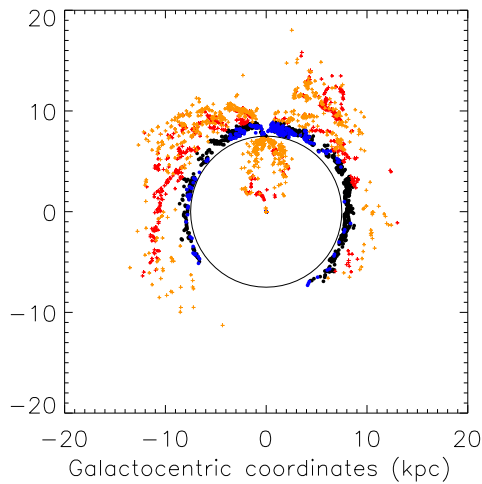}
\caption{The distribution of the gaps in the galactic plane.
The Galactic center is at the position (0,0) and the Sun at
(0, 7.5); the units are kpc. The solar circle with radius
7.5 kpc is shown. Gaps belonging to the galactic Cassini-like
division are represented by filled circles,in blue if they
are "deep" gaps, and in black if their minimum is larger than
15 K. The gaps situated outside the ring are shown with + signs, in red
if they are "deep" and in orange if they are "not very deep".}
\label{figure4}
\end{figure}

\begin{figure}
\includegraphics[width=84mm]{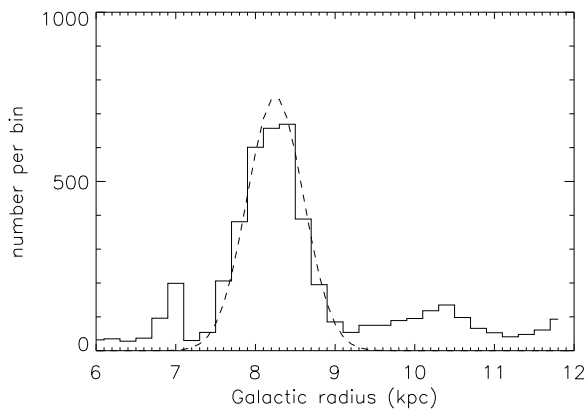}
\caption{Galactocentric distance distribution of the gaps for
 the whole  sample of measurements including different latitudes.
The Gaussian fit is centered at 8.25 kpc.}
\label{figure5}
\end{figure}

It  does not seem probable that the traditional approximation of
circular orbits adopted in calculations of kinematic distances,
instead of orbits that take into account the influence of the
galactic bar, can have any strong effect on the shape of the ring.
The part of the galactic plane which is of interest in the present
study is situated outside the solar circle. For instance, in the
study of the kinematic response of the outer stellar disk to a
central bar, M\"uhlbauer  and Dehnen  (2003) predicts that the
bar-induced radial and azimuthal motion of the LSR  should be very
small (at most a few kms$^{-1}$).

We estimated the average width of the gap by measuring the
velocities of each edge of the deep minima in the spectra.
The profiles of the minima often departs from Gaussian fits,
since there are flat minima. Instead, we just added 20 K to the
antenna temperature of the lowest point, and looked where
this threshold line intersects the edges of the line profile.
The exact value of the threshold is not important since the edges
are usually sharp. The two velocities obtained in this way were
transformed into distances from the Sun using the kinematic distance
method. The line segment defined by the two distances was then
projected onto the local radial direction.

We found an average width of 0.8 kpc, with values in the range
of 0.5 to 1.3 kpc.

The HI density in the gap can be roughly estimated with the
classical expression connecting the column density N$_H$ with
the product of antenna temperature T and velocity interval
 $\Delta$V (e.g. Vershuur, 1974): N$_H$= 1.82$\times$10$^{18}$ T $\Delta$V.
 (N$_H$  in units of cm$^{-2}$ with T in units of K and $\Delta$V
 in  kms$^{-1}$). Selecting a number of
cases for which T is smaller than 5 K over an interval of
velocity of 20 kms$^{-1}$, with corresponding length along the
line-of-sight of about 1 kpc, the resulting density is quite low,
 of the order of 0.05 cm$^{-3}$. The average density in regions
 close to the ring-like gap, in a smoothed distribution that
 does not consider the gap, is about 0.3 cm$^{-3}$ (eg. Am\^{o}res
 \&  L\'{e}pine, 2005), which means a factor 6 in density
 contrast between the gap and the surrounding regions.

\section{The gap seen with other tracers}

Although the hydrodynamical forces that produce a gas flow
that diverges from the corotation radius do not act on stars,
it is expected that the low gas density in the gap inhibits
star formation, and a lower density of young stars should
be observed. This is indeed the case, as can be seen from the
histograms of galactocentric distances of open clusters and of Cepheids
shown in Figure 6. The Open Clusters were taken from the
{\it New Catalogue of Optically visible Open Clusters and Candidates}
of Dias et al. (2002) \footnote{Available at the web page
http://astro.iag.usp.br/$_{\symbol{126}}$wilton}. We selected
from this catalog the clusters which have known  distances,
in the range of age of 2 to 40 Myr. The Cepheids were taken
from the catalog of Berdnikov et al.(2003); the complete sample
of 440 stars was used. It should be noted that the two types of
objects have their distances estimated by different methods
(main-sequence fitting and period-luminosity relation), and
different from the kinematic distance method used for HI,
so that they can be considered as independent measurements
of the gap radius.

\begin{figure}
\includegraphics[width=84mm]{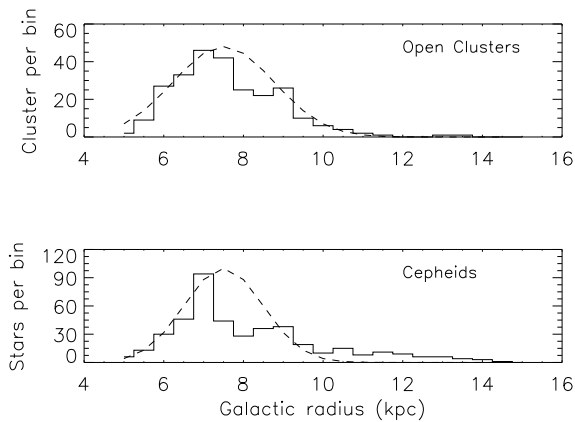}
\caption{Histograms of galactic radii of Open Clusters
and of Cepheids. The bins have a width of 0.5 kpc. Gaussian
curves centred on R$_0$ (7.5 kpc) have been drawn to emphasize
the departure from the expected symmetry with respect to the
position of the Sun}
\label{figure6}
\end{figure}

In principle, the completeness of any sample of stars decreases
with the distance from the Sun, since distant stars are more difficult
to observe. For instance, the sample of visible Mira variables
is uniformly distributed around $R_0$. This happens
because as the distance increases in the inward and outward directions,
other effects like the larger extinction towards the galactic center
and the exponential decrease of the stellar density with galactic
radius in the disk are only smooth effects. The comparison of the
histograms with symmetric curves centered on the position of the Sun
allows us to see an abrupt decrease of the density from 7.5
to 8.0 kpc. We believe that the star formation rate is almost
zero in the gap. The stars and open clusters which are formed
just outside the gap have have typical perturbation velocities
of about 10 kms$^{-1}$ with respect to the local rotation curve,
when they start their epicyclic motion around the circular
orbit (see L\'{e}pine et al. 2008 for a detailed description
of this process). As a consequence, they fill
partially the gap in about 70 Myr, half the epicycle period.
It should be noted that the Cepheids have typical ages
larger than this. It is therefore not surprising that
the gap is not seen more prominent using these two tracers.
For older stars the gap is completely smoothed out, since
the corotation resonance scatters the stellar orbits
(L\'{e}pine et al., 2003). Of course, one could interpret
the histograms of Figure 6 as showing a peak at about 7 kpc,
instead of a density decrease around 8 kpc. But at least,
the above discussion shows that there is no conflict
between the hydrogen density gap and the distribution of
young stellar tracers.

Concerning other young tracers, it is well known that H II
regions, as well as molecular clouds observed in CO,
are strongly concentrated in the spiral arms. One difficulty
is that to see clearly the ring-shaped gap we need to observe
minima along a large fraction of the gap-ring, and this is not
possible for objects which are restricted to
the spiral arms. Another problem
is how to distinguish  an inter-arm region from the ring-shaped gap.
Paladini et al. (2004), present an histogram of galactocentric
distances of HII regions, in their Figure 3. The histogram
shows a minimum at about 8.5 kpc, but this is due to a
selection effect, since the authors removed from their sample
all the objects situated along the solar circle (velocities
with modulus less than 10 km$^{-1}$). A re-analysis of the
distribution of HII regions in the Galaxy is beyond the scope
of the present work.

\section{Ring-shaped gaps in other galaxies}

Are HI ring-shaped gaps seen in other galaxies as well? In the
literature the emphasis is more often given to ring-shaped HI
emission. There are emission rings which are obviously produced by
the interaction of colliding galaxies since they present large angles
with respect to the plane of the host galaxy. However, there
are also many external rings situated
in the plane of the galactic disks. In those cases, what looks
like an external ring could be  a consequence of the
existence of a gap  that separates the  ring from the
main HI distribution. A very rough statistical analysis
can be made for instance using the unbiased  HI and optical
study of 16 nearby northern spiral galaxies by Wevers et al. (1986),
since the radial profiles of HI are given in that work.
In many cases (NGC2903, NGC3726, NGC4203, NGC4258, NGC4725,
NGC5055) the HI density profile presents a minimum which
separates a faint outer ring. This is an indication that ring-shaped
HI gaps are not rare.

The question of the connection between
a gap and  corotation is more difficult to investigate,
since the corotation radius has been determined only for a limited
number of galaxies, and there are often conflicting determinations
based on different methods. For two of the galaxies of the list given
above, there is a corotation radius estimated by Vila-Costa \&
Edmunds (1992). One is NGC2903, with R$_c$= 2.3 arcmin  while the gap,
as shown by Wevers et al. is wide and begins at about 2.8 arcmin
(but using a beamwidth smoothing of 0.5 arcmin). The other is  NGC5055,
with R$_c$=3.1 arcmin,  while the gap only starts at about 6.6 arcmin.
In this case the corotation seems to be excluded as the cause of the gap.
As examples of the gap-corotation association,
Schommer \& Sullivan (1976) states that this is clearly the case
in NGC4736. In the study of the molecular gas distribution over M83 by
Lundgren et al. (2004) we can see a ring where the gas is depleted
very near the corotation circle outside the inner arms (see eg. their
Figures 5 and 14), although the authors do not mention this feature.

\section{Consequences for the spiral structure of the Galaxy}

The existence of a "vacuum" ring is certainly able to constrain the
models of spiral arms of the Galaxy.
We argued in the introduction that there is a convergence
of the direct methods of determining the corotation radius,
telling us that corotation is close to the solar orbit.
The mean radius of the ring  is 8.3 kpc, which coincides with
the corotation radius obtained by DL, and is within the errors
of measurements of several other papers based on direct observations.
On the other hand, it is a prediction from both the classical spiral
wave theory and numerical hydrodynamic simulations that the spiral
arms produce a vacuum around corotation. In this sense, the
observation of the vacuum ring can be considered as an independent
measurement of $R_c$.
The idea of the coexisting multiple pattern speed has been
suggested by several authors (eg. Rautiainen \& Salo, H.,1999,
Minchev \& Quillen, 2006). These authors conclude that this is
a possibility in real galaxies, but do not declare that this happens
in our Galaxy. The fact that we observe only one ring-like gap,
and that this gap is a well defined one, is an argument against the
hypothesis of multiple pattern speed in the Galaxy, which would
produce multiple ring-like gaps or not well defined rings.
It should be remembered that the
self-consistent model of spiral structure proposed by AL
(1997), which seems to present multiple structures, is
based on the hypothesis of a single pattern speed. The result of AL
were confirmed by Martos et al.(2004), with the same hypotheses and
same conclusions. We avoid in the present work the discussion
of the pattern speed of the bar, which might affect the internal
regions of the Galaxy. There are arguments in favour
of a different pattern speed for the bar (eg. Dehnen, 1999), but there
is no direct measurement of it, so that we should not yet abandon
completely the hypothesis that the extremities of the bar are connected to
 the arms and both have the same rotation velocity.

Moving now to the question of  the lifetime of the
spiral pattern, what does the ring-like gap tells us about it?
In contrast with the classical theory of spiral arms,
according to  Sellwood \& Binney (2002), and  Merrifield et al. (2006),
among others, transient waves with a wide range of pattern speeds
develop in rapid succession. If the gap has an average width of
about 800 pc and the gas flow just outside the gap  has typical
velocity of the order of 0.5-1 km/s (Lacey \& Fall, 1984, Mishurov
et al., 2002, hereafter MLA), supposing that the gap
was filled with the same density of the neighbouring regions at the
initial instant, then it would have taken about 0.8 Gyr to pump out
the gas to form the gap.
This relatively short time would be the minimum age of the present
spiral structure. This is not a strong restriction; the restrictions
based on the metallicity
gradient in the disk seem to be able to impose longer lifetimes.
MLA adopted the idea that the star formation rate in the disk
is proportional to $\mid \Omega -\Omega_p \mid$, which is the velocity
of the gas with respect to the spiral arms, or the rate at which
the star-forming machine (the arms) is fed with gas. This function
presents a minimum at corotation, and is able to explain
positive slopes of metallicity beyond corotation. This interpretation
can be combined with the results of Maciel et al. (2003) which
show that the planetary nebulae younger than 4 Gyr have a metallicity
slope beyond corotation which is flat or even positive,
and very different from that of the
older PNs. This suggests that the present position
of corotation was established about 3-4 Gyrs ago, and that this is
the age of the present spiral pattern.

\section{Conclusions}

There is strong evidence in favor of the existence of a ring-like
gap in the distribution of HI in the galactic disk, similar in appearance
to the Cassini division in Saturn's rings. Although the
existence of the gap was known from the very first HI surveys,
a detailed description of it was not available up to now.
The mean radius is 8.3 kpc, for $R_0$ = 7.5 kpc, and its mean width
at half minimum of the order of 0.8 kpc. The radius of the ring
coincides with the corotation radius of the Galactic disk.
It is a prediction from both the spiral wave theory and
numerical hydrodynamic simulations that the spiral arms produce
a vacuum around corotation. The radius and width of the ring-shaped
gap coincide precisely with the results of numerical hydrodynamic
simulations performed by our group (L\'{e}pine et al., 2001)
for the same value of R$_0$. The present result is therefore
a confirmation of our correct understanding of the spiral
density waves mechanism. It tends to favour the classical idea of
a single pattern speed, since the ring-like gap is unique
and reveals a single corotation radius. It also favours a relatively
stable or long-lived spiral pattern, since in our interpretation,
the pattern must survive a time long enough to pump out the gas
from the gap.
Approximate estimates show that the HI density in the gap is about
6 times less than the mean density in the medium surrounding the
gap. Certainly this gap will have to be taken into account in
models of chemical evolution of the disk.

\section{Acknowledgements}
Eduardo Am\^ores obtained financial support for this work provided
by FAPERJ (Funda\c{c}\~{a}o Carlos Chagas Filho de Amparo \`{a}
Pesquisa do Estado do Rio de Janeiro, E-26/100.457/2008), CNPq
(Conselho Nacional para o Desenvolvimento Cient\'{i}fico e
Tecnol\'{o}gico 150772/2008-4) and Funda\c{c}\~{a}o para a
Ci\^{e}ncia e Tecnologia (FCT) under the grant
SFRH/BPD/42239/2007. We also thank the referee for the helpful
comments.

\end{document}